\documentclass[aps,
floatfix,superscriptaddress,a4paper]{revtex4}
\usepackage{graphicx,amsmath,bbm,mathrsfs,amssymb,psfrag}

\evensidemargin \oddsidemargin
\begin{document}
\author{Isiaka Aremua}
\affiliation{
Laboratoire de Physique des Mat\'eriaux et des Composants \`a
Semi-Conducteur, Universit\'e de Lom\'e (UL), B.P. 1515 Lom\'e, Togo.\\ International Chair of Mathematical Physics and Applications, ICMPA-UNESCO Chair, University of Abomey-Calavi, 072 B.P. 50 Cotonou, Republic of Benin. Email: claudisak@gmail.com}

\author{Laure Gouba}

\affiliation{
The Abdus Salam International Centre for
Theoretical Physics (ICTP),
 Strada Costiera 11,
I-34151 Trieste Italy.
Email: lgouba@ictp.it}

\title{Solving oscillations problems through affine quantization }

\date{\today}

\begin{abstract}
In this paper the benefits of affine quantization method 
are highlighted through oscillation problems. We show how 
affine quantization is able to solve oscillation problems 
where canonical quantization fails.
\end{abstract}

\maketitle

\section{Introduction}\label{sec1}
Quantization is generally understood as a correspondence between a classical and a quantum theory and it appears natural in a quantization procedure that classical variables be chosen before choosing what quantum operators to employ since classical physics that refers to pre-1900 physics is seen as the oldest theory. How to formulate a quantum theory if a classical system is given? The response to that question is addressed in some overviews of the better known quantization techniques found in the current literature and used both by physicists and mathematicians \cite{ali1,ali2}.
However, quantum mechanics, like any other physical theory, classical mechanics, electrodynamics, relativity, thermodynamics, cannot be derived. The laws of quantum mechanics, expressed in mathematical form, are the results of deep physical intuition, as indeed, are all other physical theories. Their validity can only be checked experimentally. From this point of view, quantization is not a method for deriving quantum mechanics, rather it is a way to understand the deeper physical reality which underlies the structure of both classical and quantum mechanics. In his most recent paper J. R. Klauder has addressed how to overcome the two separate languages and create a smooth and common procedure that provides a clear and continuous passage between the conventional distinction of either a strictly classical and quantum expressions \cite{klau1}. A connection between the classical and the quantum realms is provided by the coherent states \cite{klau,gaz, twaangaz} and in that sense we consider the procedure of coherent states quantization and integral quantization among the new trends of quantization procedures \cite{gaz1,gaz2,gaz3,gaz4}. We have been particularly interested in new developments of quantization techniques, \cite{aremi1, aremi2, lau1, lau2}, bearing in our mind that affine coherent states framework is useful for investigating quantum gravity
\cite{jrk4}. 

The main motivation of investigating in affine quantization is that although Dirac's canonical quantization works reasonably well and has got success it has some severe shortcoming from a theoretical point of view when it comes to non-trivial spaces.
 We are interested for instance in examining quantization procedure on configuration spaces other than $\mathbb{R}^2$.
 The simple example of a particle restricted to move on the positive real line is a well illustration.  In such case, the configuration space is $\mathbb{Q} = \mathbb{R}^+$. At a classical level it is reasonable to use the position $x$ and the momentum $p_x$ as classical observables and they satisfy the usual Poisson brackets relations. Promoting naively the classical variables respectively to operators 
 $\hat x \equiv x$ and $\hat p_x \equiv -i\hbar\frac{\partial}{\partial_x}$, it turns out that the momentum operator $\hat p_x$ is not self adjoint on the Hilbert space $\mathcal{H} = L^2(\mathbb{R}^+, dx)$ and this means that a straightforward application of Dirac's canonical quantization method is impossible. The method of affine quantization expands similar procedures of canonical quantization to solve problems that canonical quantization can not solve \cite{jrk1,jrk2,jrk3, frion, fanuel}. For instance in a recent work, the usefullness of the affine quantization has been highlighted in the quantization of the classical Brans-Dicke Theory \cite{frion}.  Indeed, the authors have used affine quantization rather than canonical quantization, since the domain of the variables involved (scale factor and scalar field) is the real half-line and the corresponding phase space can be identified with the affine group. 
 
In order to make affine quantization procedure our favorite to be solving more complex problems that canonical quantization cannot solve such as gravity, we start with some simple one-dimensional models. Indeed, the solution of one-dimensional (1D) problems is usually the first approach to the calculation of quantum effects in physics, because solutions of the Schr\"odinger equation in one dimension are technically easier to find. Although considered as simple, a careful look into one-dimensional problems shows that they cannot always be solved without serious challenges. One of us has recently examined the affine quantization on the half line where the case of the free particle and the harmonic oscillator have been solved \cite{lg1}. The spectrum of the half harmonic oscillator is given in terms confluent hypergeometric functions 
 \begin{equation}\label{eqintro}
 \phi_n(x) = \sqrt{2(n+1)}(\frac{m\omega}{\hbar}) x^{3/2}e^{-\frac{m\omega}{2\hbar} x^2} {}_1F_1(-n,2,\frac{m\omega}{\hbar} x^2), \quad E_n = 2(n+1)\hbar\omega,\: n= 1,2,\ldots
 \end{equation}
 As we can see the energy eigenvalues are equally spaced as the one of the linear harmonic oscillator and the eigenfunctions are the well known solution of the harmonic oscillators with repulsive potential proportional to $1/x^2$. The polynomial confluent hypergeometric functions are actually proportional to associated Laguerre polynomials with quadratic variable and the latter are proportional to Hermite polynomial when the parameter is $1/2$, more details on the links between those special functions can be found in the literature
 \cite{almeida, bergeron, grith}.  
Solving the case of the half harmonic oscillator brought to  
our attention the case of a coupled half harmonic oscillators. The motivation to study this specific system is that the decoupling gives different scenario than the case of a coupled two full harmonic oscillator and a different spectrum. An other benefit of the affine quantization is the possibility to recover the solution of the full harmonic oscillator by fixing the symmetry point and changing the end point with the aim to recover the solution of the full harmonic oscillator. That scheme is to go from affine quantization to canonical quantization, the reverse scheme not being possible.
 In section (\ref{sec2}), we study the interaction of two half harmonic oscillators, in section (\ref{sec3}) we examine how to restore the solutions of the full harmonic oscillator from the solutions of the half harmonic oscillator. Concluding remarks are given in section (\ref{sec4}).
 \section{Interactions between two half harmonic oscillators}\label{sec2}
 We consider the two harmonic oscillators represented 
by the classical Hamiltonian on the half line
\begin{equation}\label{eq1}
H_c (x_1,x_2,p_{x_1},p_{x_2}) = \frac{1}{2m}p_{x_1}^2 + \frac{1}{2m}p_{x_2}^2 
+ \frac{1}{2}mw^2x_1^2 + \frac{1}{2}mw^2x_2^2 + gx_1x_2,
\end{equation}
where $(x_1,x_2)\in \mathbb{R}^+\times\mathbb{R}^+$ and 
$(p_{x_1},p_{x_2})\in \mathbb{R}\times \mathbb{R}$, $g$ is a coupling constant that we impose to bounded as $|g| < m\omega^2$. The associated 
nonvanishing Poisson brackets are given by
\begin{equation}\label{eq2}
\{x_1,p_{x_1}\} = 1;\quad \{x_2,p_{x_2}\} = 1.
\end{equation}
Let's redefine the model by mean of change of variables as follows
\begin{equation}\label{eq3}
y_1  = x_1 + x_2;\quad y_2  = x_1 -x_2; \quad
p_{y_1} = p_{x_1} + p_{x_2};\quad p_{y_2} = p_{x_1}-p_{x_2},
\end{equation}
where $(y_1,y_2)\in \mathbb{R}^+ \times \mathbb{R}$ and 
$(p_{y_1}, p_{y_2})\in \mathbb{R}\times \mathbb{R}$.
Regarding the change of variables in equation (\ref{eq3}), 
the Hamiltonian in equation (\ref{eq1}) is rewritten as 
\begin{equation}\label{eq4}
\tilde{H}_c = \frac{1}{4m}p_{y_1}^2 +\frac{1}{4m}p_{y_2}^2
+ \frac{1}{4}(m\omega^2 + g)y_1^2 + \frac{1}{4}(m\omega^2 - g)y_2^2.
\end{equation}
The Hamiltonian in equation (\ref{eq4}) is now describing 
a model of two independent harmonic oscillators on 
$\mathbb{R}^+\times \mathbb{R}$, the fundamental variables being 
$y_1,\;y_2,\;p_{y_1},\; p_{y_2}$ from which the associated nonvanishing Poisson brackets are given by 
\begin{equation}\label{eq5}
\{y_1,p_{y_1}\} = 2;\quad \{y_2,p_{y_2}\} = 2.
\end{equation}
We may perform some rescaling to have the kind of usual 
Poisson brackets structure in case it is necessary.
Since $y_1 > 0$, a canonical quantization procedure may fail 
as the associated conjugate momenta $p_{y_1}$ fails to be self adjoint. An alternative is to perform affine quantization procedure. In order to do that we consider the dilation variable $d_{y_1} = p_{y_1} y_1$ which together with $y_1$ form the fundamental Poisson bracket $\{y_1, d_{y_1}\} = 2y_1$, they are not canonical coordinates
but form a Lie algebra and are worthy of consideration as new pair of classical variables. The fundamental variables are now $y_1,\;y_2,\;d_{y_1},\;p_{y_2}$. The classical hamiltonian (\ref{eq4}) and the equation (\ref{eq5}) are now substituted by 
\begin{equation}
\tilde{H}_c = \frac{1}{4m}d_{y_1}(y_1^{-2})d_{y_1} +\frac{1}{4m}p_{y_2}^2 + \frac{1}{4}(m\omega^2 + g)y_1^2 + \frac{1}{4}(m\omega^2 - g)y_2^2,
\end{equation}
and
\begin{equation}
\{y_1,d_{y_1}\} = 2y_1;\quad \{y_2,p_{y_2}\} = 2 .
\end{equation}
For the quantization procedure, the classical variables 
are promoted to operators as follows

$y_1 \rightarrow \hat{y}_1$; 
$y_2\rightarrow \hat{y}_2$;
$d_{y_1}\rightarrow \hat{d}_{y_1}$;
$ p_{y_2}\rightarrow \hat{p}_{y_2}$,
and 
\begin{equation}\label{eq6}
\tilde{H}_q =  \frac{1}{4m}\hat{d}_{y_1} \hat{y_1}^{-2} \hat{d}_{y_1} +\frac{1}{4m}\hat{p}_{y_2}^2 
+ \frac{1}{4}(m\omega^2 + g)\hat{y}_1^2 + \frac{1}{4}(m\omega^2 - g)\hat{y}_2^2.
\end{equation}
The nonvanishing commutators are 
\begin{equation}
[ \hat{y}_1,\hat{d}_{y_1} ] = 2i\hbar\hat{y}_1;\quad 
[\hat{y_2},\hat{p}_{y_2} ] = 2i\hbar .
\end{equation}
The Schr\"odinger representation is as follows
\begin{eqnarray}
\hat{y}_1\psi(y_1,y_2) &=& y_1\psi(y_1,y_2);\\
\hat{y}_2\psi(y_1,y_2) &=& y_2\psi(y_1,y_2);\\
\hat{d}_{y_1}\psi(y_1,y_2) &=& -i\hbar(y_1\partial_{y_1} +\frac{1}{2})\psi(y_1,y_2);\\
\hat{p}_{y_2}\psi(y_1,y_2) &=& -i\hbar\frac{\partial}{\partial_{y_2}}\psi(y_1,y_2).
\end{eqnarray}
In order to solve the Schr\"odinger equation, 
\begin{equation}
i\hbar\partial_t\psi(y_1,y_2,t) = H_q\psi(y_1,y_2,t),
\end{equation}
we assume now that the constant of coupling $g$ is bounded as
$ 0 < g < m\omega^2 $.
In presence of autonomous system, we consider the Ansatz
\begin{equation}
\psi(y_1,y_2,t) = e^{-itE_{y_1,y_2}/\hbar}\phi(y_1,y_2),
\end{equation} 
the corresponding time-independent eigenvalue equation is 
\begin{equation}\label{eq7}
H_q(\hat y_1,\hat y_2, \hat d_{y_1},\hat p_{y_2})\phi(y_1,y_2)
= E_{y_1,y_2}\phi(y_1,y_2).
\end{equation}
From the expression of equation (\ref{eq6}), the equation (\ref{eq7}) can be decoupled and in that sense we may view 
$\phi(y_1,y_2)$ as $\phi(y_1,y_2) = \phi(y_1)\phi(y_2)$ 
and $\phi(y_1)$ and $\phi(y_2)$ being determined separately, 
associated respectively to $E_{y_1}$ and $E_{y_2}$ with 
$E_{y_1,y_2} = E_{y_1} + E_{y_2}$.
We have then to solve 
\begin{equation}\label{eqo1}
\left[ -\frac{d^2}{dy_1^2} + \frac{3}{4}\frac{1}{y_1^2} 
+ \frac{m}{\hbar^2}(m\omega^2 + g)y^2_1\right]\phi(y_1) 
= \frac{4mE_{y_1}}{\hbar^2}\phi(y_1),
\end{equation}
and 
\begin{equation}\label{eqo2}
\left[-\frac{d^2}{dy_2^2} + \frac{m}{\hbar^2}(m\omega^2 -g)y_2^2\right]\phi(y_2) = \frac{4mE_{y_2}}{\hbar^2}\phi(y_2).
\end{equation}
It is easy to derive the solutions of (\ref{eqo1}) from \cite{lg1} as follows 
\begin{equation}\label{hypf1}
\phi_n(y_1) = \sqrt{2(n+1)} (\frac{m\omega}{\hbar}\sqrt{1 + g/(m\omega^2)})y_1^{3/2}e^{-\frac{m\omega}{2\hbar}\sqrt{1 + g/(m\omega^2)}y_1^2}
\mbox{}_1F_1\left(-n, 2, \frac{m\omega}{\hbar}\sqrt{1 + g/(m\omega^2)}y_1^2 \right).
\end{equation}
Setting $\alpha_1 = \frac{m\omega}{\hbar}\sqrt{1+g/(m\omega^2)}$, we have 
\begin{equation}
\phi_n(y_1) = \sqrt{2(n+1)}\alpha_1y_1^{3/2}e^{-\frac{\alpha_1}{2} y_1^2}
\mbox{}_1F_1\left(-n, 2,\alpha_1 y_1^2\right),
\end{equation}
\begin{equation}\label{hypen1}
E_{y_1,n} = (n+1)\hbar\omega\sqrt{1 + g/(m\omega^2)},\; n= 1, 2, \ldots
\end{equation}
The solutions of (\ref{eqo2}) is also easy to derive from the 
one of simple harmonic oscillator as
\begin{equation}
\phi_n(y_2) = \left( \frac{\sqrt{\alpha_2}}{2^n n!\sqrt \pi}\right)^{1/2}  e^{-\alpha_2 y_2^2/2}\:H_n(\sqrt{\alpha_2} y_2),\quad 
\alpha_2 =  \frac{m\omega}{\hbar}\sqrt{1-g/(m\omega^2)},
\end{equation}
where $H_n$ represents the Hermite polynomial of degree $n$ and the 
eigenvalues are given by the simple formula 
\begin{equation}
E_{y_2,n} = (n+1/2)\frac{\hbar\omega}{2}\sqrt{1 - g/(m\omega^2)}.
\end{equation}
\section{From affine quantization to canonical quantization}\label{sec3}
The solution of the half harmonic oscillator solved in 
(\cite{lg1}) is displayed in equation (\ref{eqintro}).
We have been curious on the possibility of obtaining the solution of the full harmonic oscillator from the equation (\ref{eqintro}), that is getting back to canonical quantization from affine quantization. The question is whether it is possible to recover the solutions of the usual harmonic oscillator on the line from the solutions of the half harmonic oscillator?
In order to analyse that question, we consider a positive real $b\ge 0$, where $-b < x$. In the situation of the half line, the end point and the symmetry point coincide to $0$. Now we imagine the situation in which we save the symmetry point that remains $0$ while the end point can move toward negative infinity. The situation in the quantum picture is described by the Hamiltonian 
\begin{equation}
\hat H_a(\hat x, \hat d_x) = \frac{1}{2m}\left(
(\hat d_x +b\hat p_x)(\hat x +b)^{-2}(\hat d_x +b\hat p_x)\right)
+ \frac{1}{2}m\omega\hat x^2, 
\end{equation}
 where $\hat d_x$ is the dilation operator defined in the previous sections and 
 \begin{equation}
 [\hat x, \hat p_x] = i\hbar, \quad
 [\hat x, \hat d_x] = i\hbar \hat x\;.
 \end{equation}
The corresponding time independent Schr\"odinger equation is given by 
\begin{equation}\label{hext1}
\left[ -\frac{d^2}{dx^2} + \frac{3}{4}\frac{1}{(x+b)^2} + \frac{m^2\omega^2}{\hbar^2} x^2 \right] \phi(x) = \frac{2mE}{\hbar^2}\phi(x),
\end{equation} 
When $b = 0$, we recover the problem in Section 3 of \cite{lg1} brieffly introduced in section (\ref{sec1}) and 
when $b\rightarrow +\infty$ the equation (\ref{hext1}) tends to the one of the harmonic oscillator on the line.
Indeed for $b$ sufficiently large,  we assume $\vert \frac{x}{b}\vert < 1$, and therefore the approximation 
\begin{equation}\label{bexpansion}
\frac{1}{(x+b)^2}\sim \frac{1}{b^2}\left(
1 -\frac{2x}{b} +3\frac{x^2}{b^2} -4\frac{x^3}{b^3}
+ 5\frac{x^4}{b^4} + \ldots \right)
\end{equation}
If the expansion is considered up to the fourth level, then, 
inserting the equation (\ref{bexpansion}) in the equation
(\ref{hext1}), the potential to consider is 
\begin{equation}\label{fpotential}
V(x) = \frac{3}{4b^2}\left(
1 -\frac{2x}{b} +3\frac{x^2}{b^2} -4\frac{x^3}{b^3}
+ 5\frac{x^4}{b^4} \right) + 
\frac{m^2\omega^2}{\hbar^2} x^2.
\end{equation}
 The potential in equation (\ref{fpotential}) converges strongly as 
$b \rightarrow \infty$ to the one of the full harmonic oscillator. 
While the equation (\ref{hext1}) is not obvious to solve analytically, we guess that the limit  $b\rightarrow +\infty$ leads from affine to canonical and should lead to usual even and odd about the symmetry point $x=0$ eigenfunctions.
A temptative way to solve quasi-analytically the equation (\ref{hext1}) is to consider the potential in the equation (\ref{fpotential}) and  
consider the method of resolution by Christiane Quesne in her paper titled `` Quasi-exactly solvable polynomial extensions of the quantum harmonic oscillator " \cite{cquesne}.
The study of this problem is another example of how affine quantization deals with harmonic oscillator problems that canonical quantization cannot resolve.
\section{Concluding remarks}\label{sec4}
The coupling of two half harmonic oscillator lead to 
a problem of a decoupled harmonic oscillator on the full 
line and on the half line. In such situation, we show that canonical quantization procedure and affine quantization procedure interviened simultaneously in elegant way. We also show the possibility of getting back to canonical quantization from affine quantization.
Our investigations on the oscillations problems on non trivial configurations space provided an insight to investigate affine coherent states and their applications y include solving more complicated problems like how to quantize gravity using affine quantization techniques. 

{\bf Acknowledgements }: I. Aremua and L. Gouba would like to gratefully thank Professor J. R. Klauder for drawing the consequent oscillators problems to their attention and for the useful instructions.

\end{document}